\documentclass[aps,prl,reprint,superscriptaddress]{revtex4-2}
\usepackage[utf8]{inputenc}
\usepackage{natbib}

\usepackage{amsmath}
\usepackage[toc,page]{appendix}
\usepackage{bm}
\usepackage{ amssymb }
\usepackage{graphicx}
\usepackage {epstopdf}
\usepackage{csquotes}
\usepackage{float}
 \usepackage{color, soul}

\begin{document}

\title{Half-integer conductance plateau at the $\nu = 2/3$ fractional quantum Hall state in a quantum point contact}

\author{J. Nakamura}
\email[]{jnakamur@purdue.edu}

\affiliation{Department of Physics and Astronomy, Purdue University}
\affiliation{Birck Nanotechnology Center, Purdue University}

\author{S. Liang}
\affiliation{Department of Physics and Astronomy, Purdue University}
\affiliation{Birck Nanotechnology Center, Purdue University}

\author{G. C. Gardner}
\affiliation{Birck Nanotechnology Center, Purdue University}
\affiliation{Microsoft Quantum Lab West Lafayette}

\author{M. J. Manfra}
\email[]{mmanfra@purdue.edu}
\affiliation{Department of Physics and Astronomy, Purdue University}
\affiliation{Birck Nanotechnology Center, Purdue University}
\affiliation{Microsoft Quantum Lab West Lafayette}
\affiliation{Elmore Family School of Electrical and Computer Engineering, Purdue University}
\affiliation{School of Materials Engineering, Purdue University}

\date{\today}

\begin{abstract}
The $\nu = 2/3$ fractional quantum Hall state is the hole-conjugate state to the primary Laughlin $\nu = 1/3$ state. We investigate transmission of edge states through quantum point contacts fabricated on a GaAs/AlGaAs heterostructure designed to have a sharp confining potential. When a small but finite bias is applied, we observe an intermediate conductance plateau with $G = 0.5 \frac{e^2}{h}$. This plateau is observed in multiple QPCs, and persists over a significant range of magnetic field, gate voltage, and source-drain bias, making it a robust feature. Using a simple model which considers scattering and equilibration between counterflowing charged edge modes, we find this half-integer quantized plateau to be consistent with full reflection of an inner counterpropagating -1/3 edge mode while the outer integer mode is fully transmitted. In a QPC fabricated on a different heterostructure which has a softer confining potential, we instead observe an intermediate conductance plateau at $G = \frac{1}{3} \frac{e^2}{h}$. These results provide support for a model at $\nu = 2/3$ in which the edge transitions from a structure having an inner upstream -1/3 charge mode and outer downstream integer mode to a structure with two downstream 1/3 charge modes when the confining potential is tuned from sharp to soft and disorder prevails. 

\end{abstract}
\date{\today}

\maketitle

The quantum Hall effect occurs when a two-dimensional electron gas (2DEG) is cooled to low temperature and placed in a strong magnetic field. Fractional quantum Hall states \cite{Tsui1982, Laughlin1983} exhibit fractionally quantized Hall conductance, and have excitations that behave as fractionally charged quasiparticles which obey anyonic braiding statistics \cite{Leinaas1977, Wilczek1982, Halperin1984, Arovas1984,Bartolomei2020,Nakamura2020}.

The concept of particle-hole symmetry plays a central role in the hierarchical construction of daughter states from the $\nu=1/m$ Laughlin series \cite{Haldane1983, Girvin1984, Halperin1984}. If Landau-level mixing is neglected, particle-hole symmetry is expected for each Landau level \cite{Girvin1984}; a similar approximate particle-hole symmetry exists for composite fermions \cite{Jain2017}. The $\nu = 2/3$ fractional quantum Hall state can be considered to be the hole-conjugate state to the $\nu = 1/3$ state \cite{Girvin1984, Jain2017}. In this picture, the fully filled $\nu = 1$ state is considered to be the vacuum, and holes form a $\nu = 1/3$ state on top of this vacuum, leading to an overall 2/3 filling factor.  The 2/3 state has been investigated in numerous theoretical works \cite{Macdonald1990, Jain2012, Hu2008}.

The edge mode structure in hierarchical states may possess several branches \cite{Macdonald1990, Wen1991}. Moreover, multiple edge structures have been proposed for the 2/3 state. The edge structure which most straightforwardly follows from the hole-conjugate nature of the state was proposed by MacDonald \cite{Macdonald1990, Macdonald1991}. In this picture, the edge consists of an outer downstream integer edge mode (from the underlying $\nu = 1$ state) and an inner, counterpropogating -1/3 edge mode. In this context, counterpropagating (or upstream) edge modes travel in the direction opposite to the conventional quantum Hall chiral edge current (edge modes following the conventional direction may be referred to as downstream). Equilibration between the two edge modes leads to a total conductance of $\frac{2}{3} \frac{e^2}{h}$. A different edge structure was proposed by Kane and collaborators \cite{Kane1994} in which disorder dominates and drives an edge phase transition to a single downstream charge mode with conductance $\frac{2}{3}\frac{e^2}{h}$ and a counterpropagating neutral mode. A model developed by Meir and coworkers considered the impact of soft confining potential in addition to disorder and predicted a reconstructed edge involving the formation of a strip with filling factor $\nu = 1/3$ at the edge leading to two downstream charge modes with conductance $\frac{1}{3} \frac{e^2}{h}$ along with counterpropagating neutral modes \cite{Meir1993, Meir2013}. Other theoretical works have also examined the possibility of incompressible strips with fractional fillings when the confining potential is soft \cite{Chklovskii1992, Beenakker1990, Chang1990}, including recent works investigating $\nu = 1$ \cite{Khanna2021} and $\nu = 1/3$ \cite{Khanna2022}. Experimental studies of the $\nu = 2/3$ state in quantum point contacts have usually shown an intermediate conductance plateau with $G = \frac{1}{3} \frac{e^2}{h}$ when the QPC gates are negatively biased to bring the edges close together, which supports the existence of the Meir edge structure in those experiments \cite{Heiblum2009, Ensslin2014, Heiblum2017}. Experiments in which this $G = \frac{1}{3} \frac{e^2}{h}$ intermediate plateau was observed also detected upstream noise attributed to counterpropagating neutral modes \cite{Heiblum2010, Heiblum2017}. These neutral modes are likely to be detrimental to interference experiments because their entanglement with the charge modes leads to dephasing \cite{Gefen2016, Heiblum2019-2}. 

\begin{figure}[h]
\def\ffile{Device.pdf}
\centering
\includegraphics[width=1.0\linewidth]{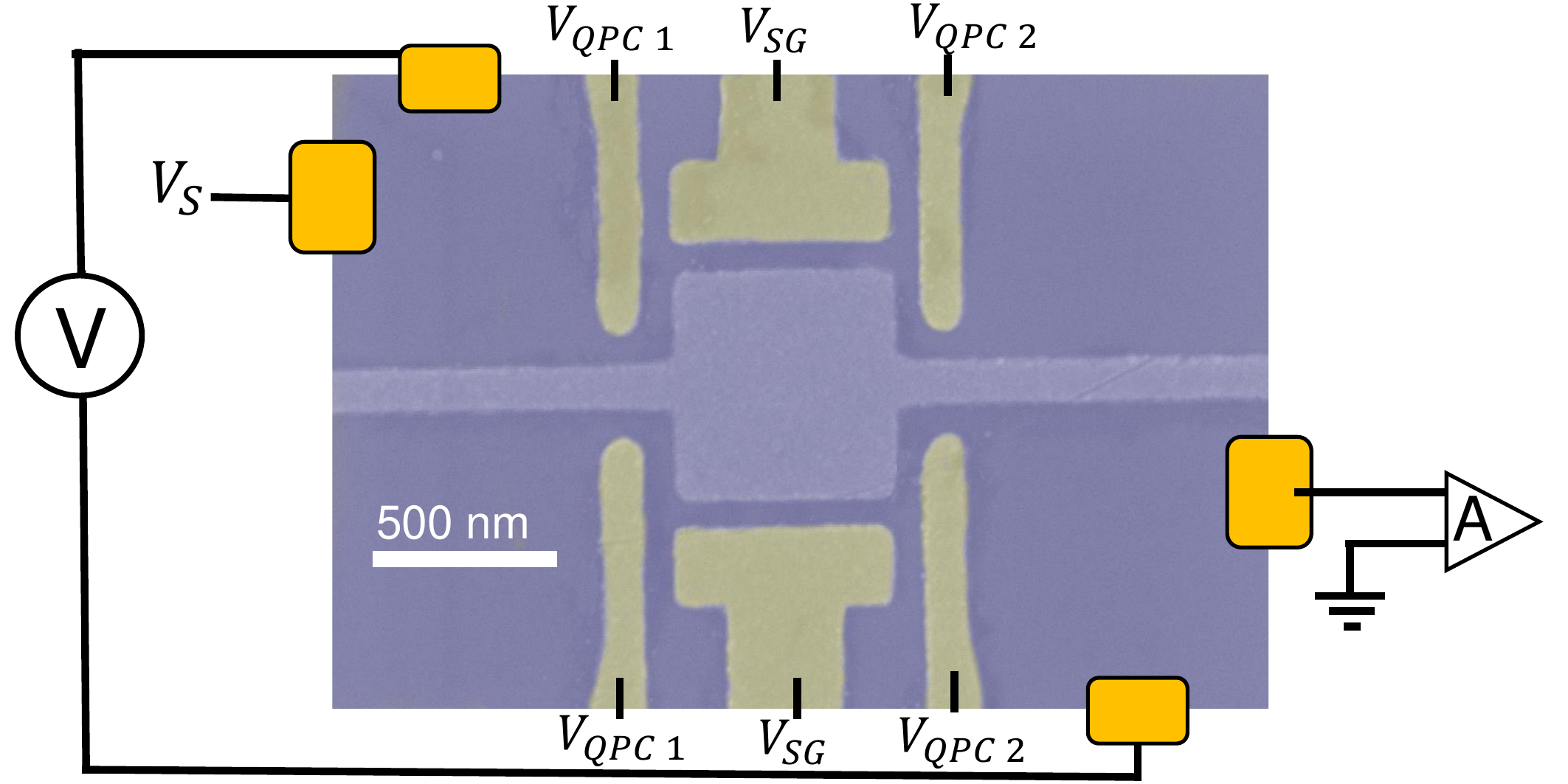}
\caption{\label{Integer_data}False-color SEM image of device with two QPCs.}
\end{figure}

The device investigated in our experiments consists of two quantum point contacts (QPCs) separated by a pair of side gates, and utilizes a GaAs/AlGaAs heterostructure with auxiliary screening wells above and below the main quantum well (shown in Supp. Fig. 1 and described in Supp. Section 1). This geometry can be used to define a Fabry-Perot interferometer \cite{Chamon1997, Halperin2011, Halperin2021, Halperin2022, Nakamura2019, Nakamura2020, Nakamura2022}, but in this work we probe the QPCs independently without inducing interference. A false color SEM image is shown in Fig. \ref{Integer_data}a. Each QPC has a separation of 300 nm. We probe the QPCs individually, measuring the conductance of one QPC while tuning the gate voltage on the other to full transmission. We apply a voltage $V_{SD}$ on the source contact and measure current at the drain. The voltage drop across the device is measured, as illustrated in Fig. \ref{Integer_data}, and the conductance as $G = \frac{I}{V}$ is calculated. The semiconductor heterostructure employed here has been designed to produce sharp confining potentials consistent with our recent interferometry experiments \cite{Nakamura2019, Nakamura2020, Nakamura2022}. 

\begin{figure*}[t]
\def\ffile{nu_2_3_QPCs_V3.pdf}
\centering
\includegraphics[width=1.0\linewidth]{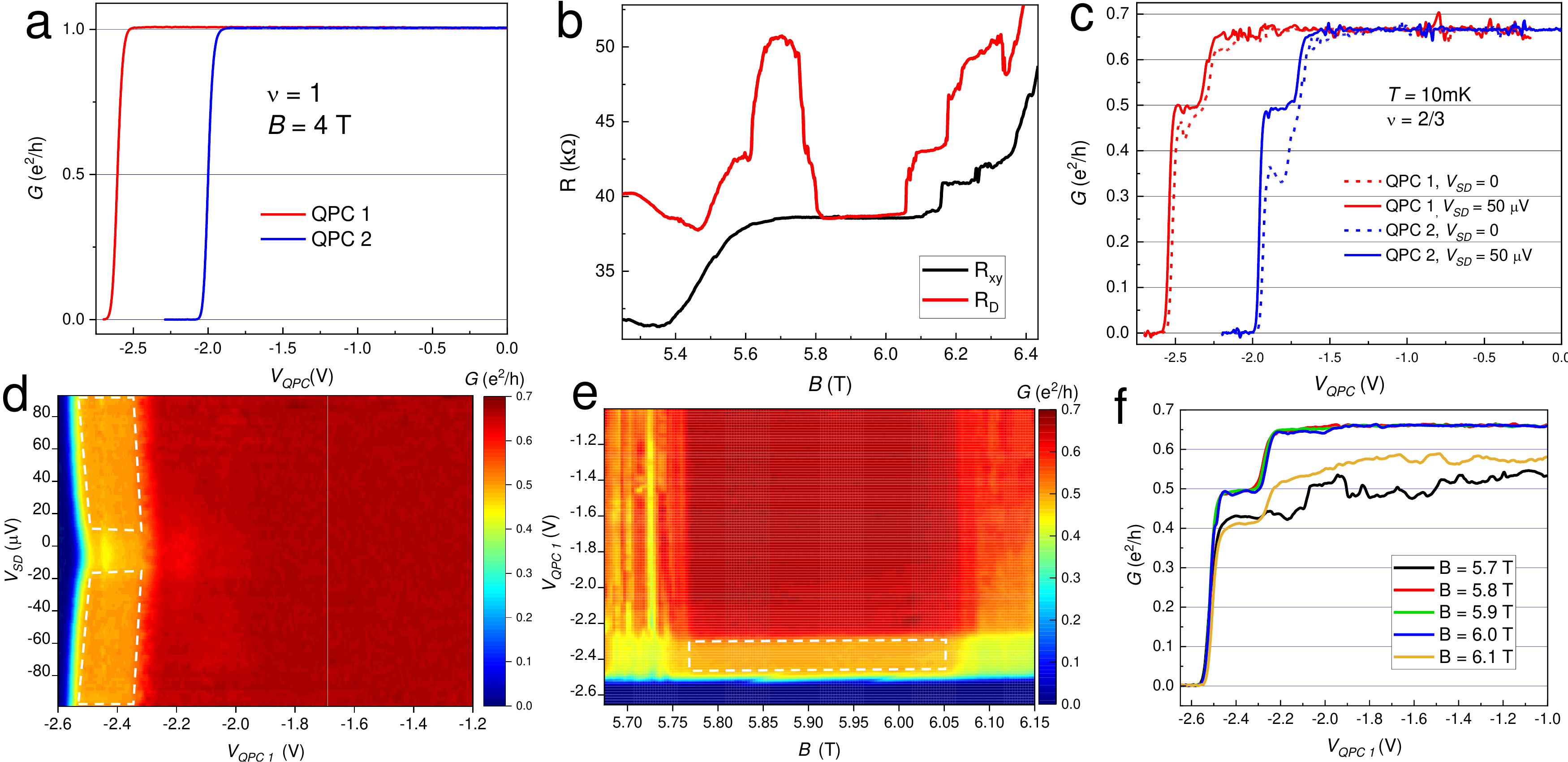}
\caption{\label{nu=2_3_QPCs}a) QPC Sweeps at $\nu = 1$. QPC 2 is shifted by 0.5 V for clarity. b) Bulk transport $R_{xy}$ and $R_D$ measured across the device. For the $R_D$ measurement, both QPCs and the side gates are set to -0.3 V so that the 2DEG under the gates is depleted, but minimal backscattering occurs in the QPCs. c) QPC sweeps at $\nu = 2/3$ with DC source-drain bias of 0 V (dashed lines) and 50 $\mu$V (solid lines). With 50 $\mu$V bias, an intermediate plateau at 0.5 $e^2/h$ appears in both QPCs. d) Differential conductance for QPC 1 as a function of QPC gate voltage and source-drain bias $V_{SD}$. White dashed lines indicate the regions where $G \approx 0.5 \frac{e^2}{h}$. e) Differential conductance for QPC 1 versus magnetic field and QPC voltage with $V_{SD} = 50$ $\mu$V. f) Line cuts of conductance versus QPC voltage at different magnetic fields.}
\end{figure*}

In Fig. \ref{nu=2_3_QPCs}a we show conductance versus gate voltage at the $\nu = 1$ integer quantum Hall state. The QPCs exhibit a single wide plateau (corresponding to full transmission of the $\nu = 1$ edge state) and a very abrupt drop in conductance when the edge state is pinched off in the QPC. This abrupt pinchoff suggests that the confining potential is sharp, as expected for devices utilizing the screening well structure \cite{Harshad2018}, since minimal backscattering occurs until the edges are brought very close together. This behavior contrasts with the behavior generally observed in QPCs on standard GaAs heterostructures without screening wells at $\nu = 1$, which often show gradual decreases in conductance, resonances, and sometimes intermediate fractionally quantized plateaus due to edge reconstruction \cite{Heiblum2019-2, Ensslin2014}.

Next we operate the device at bulk filling factor $\nu = 2/3$. Fig. \ref{nu=2_3_QPCs}b shows the bulk Hall resistance $R_{xy}$ and the diagonal resistance $R_D$ measured across the device with the 2DEG under the gates depleted, but with small enough voltages on the gates that minimal backscattering occurs in the QPCs, $V_{QPC 1}=V_{QPC 2} = -0.3$ V. A wide plateau is visible in $R_{xy}$ with resistance $\frac{3}{2} \frac{h}{e^2}\approx 38.7$ k$\Omega$, corresponding to the $\nu = 2/3$ state, and $R_D$ exhibits a nearly quantized plateau over a somewhat narrower region. Fig. \ref{nu=2_3_QPCs}c shows conductance versus gate voltage for both QPCs at $B$ = 5.9 T, near the center of the plateau. Between 0 and -2 V both QPCs show conductance close to $\frac{2}{3} \frac{e^2}{h}$, indicating close to full transmission of all edge states. When there is zero applied DC bias (dashed lines), the QPCs exhibit exhibit a somewhat less sharp pinchoff than at $\nu = 1$, but unlike the experiments in Ref. \cite{Heiblum2009, Heiblum2017}, there is no intermediate quantized conductance plateau at $G=\frac{1}{3} \frac{e^2}{h}$. The behavior changes when a finite source-drain bias of 50 $\mu$V is applied, shown in the solid lines; in this situation the applied source-drain bias is a combination of a 10 $\mu$V AC bias which is used to probe the differential conductance and the 50 $\mu$V DC bias. With the finite bias both QPCs exhibit an intermediate conductance plateau; however, the value of the intermediate conductance plateaus is 0.5$\frac{e^2}{h}$ rather than the $\frac{1}{3} \frac{e^2}{h}$ seen in previous experiments \cite{Heiblum2009, Heiblum2017} (the quantization is not exact, but the conductance is within 2\% of 0.5 $\frac{e^2}{h}$ for both QPCs). This observation suggests that a different edge structure may be present in our system, although it is not obvious why the intermediate plateau only occurs when finite bias is applied. Intermediate conductance plateaus are generally interpreted as occurring when one or more inner edge states are fully reflected while the outer edge states are fully transmitted, leading to some range of gate voltage before the next outer edge state starts to be backscattered when there is an incompressible region in the middle of the QPC and the conductance is constant.  

In Fig. \ref{nu=2_3_QPCs}d we plot the differential conductance for QPC 1 as a function of $V_{SD}$ and QPC voltage; dashed lines indicate regions where $G \approx 0.5 \frac{e^2}{h}$. The 0.5 $\frac{e^2}{h}$ plateau develops as finite bias is applied, and the effect is nearly symmetric around zero bias. The value of the intermediate plateau does not change significantly when $V_{SD}$ is increased past $\approx 25$ $\mu$V. The intermediate plateau persists over a significant range of magnetic field as seen in Fig. \ref{nu=2_3_QPCs}e and f. We have also observed this $G = 0.5 \frac{e^2}{h}$ conductance plateau in another set of QPCs on a different chip fabricated on the same wafer (Supp. Fig. 3).

\begin{figure}[t]
\def\ffile{nu_2_3_cartoons_v5.pdf}
\centering
\includegraphics[width=1.0\linewidth]{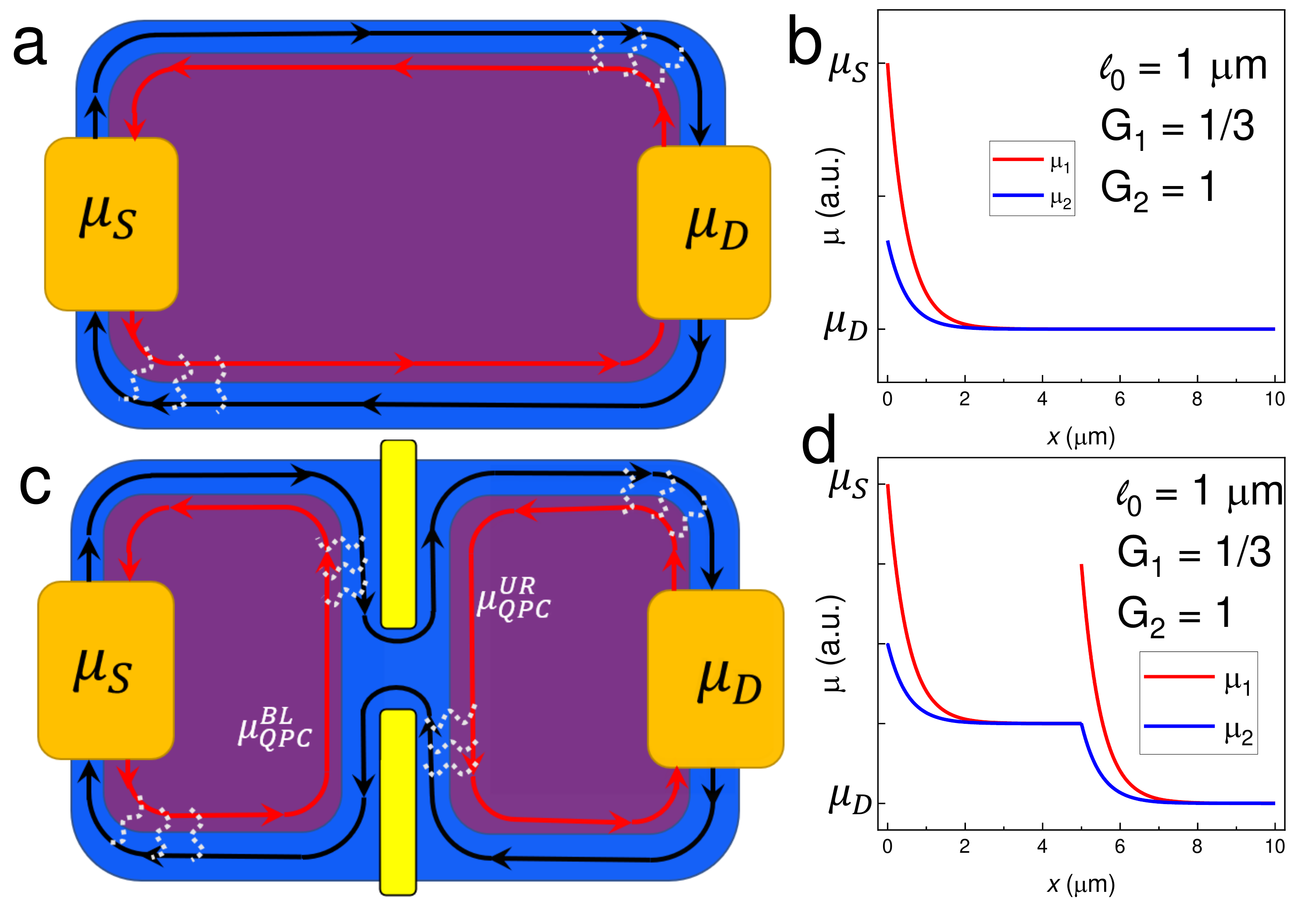}
\caption{\label{nu=2_3_cartoons}a) Schematic of edge states in a Hall bar in the Macdonald picture, with a counterpropogating inner -1/3 edge state (red arrows) and outer integer edge state (black arrows). Dashed white lines indicate equilibration near the Ohmic contacts. b) Qualitative plot of the chemical potential for the inner mode (red) and outer mode (blue) on the bottom edge of the Hall bar. c) Schematic of a QPC which fully transmits the outer edge state and fully reflects the inner one. There will be additional equilibration in the vicinity of the QPC. d) Chemical potential versus position for the Hall bar with QPC and full reflection of the inner mode. }
\end{figure}

The 0.5 $\frac{e^2}{h}$ conductance plateau is not straightforward to interpret, since no proposed edge structure contains an outer edge state with $\Delta \nu = 0.5$ which would carry 0.5 $\frac{e^2}{h}$ on its own (unlike the $\nu = 5/2$ fractional state, $\nu = 1/2$ is expected to be a compressible sea of composite fermions that does not form a quantum Hall state \cite{Halperin1993}). However, a possible explanation for the 0.5 $\frac{e^2}{h}$ plateau could come from the counterpropagating charge mode as proposed in the Macdonald edge structure \cite{Macdonald1990, Macdonald1991}. Following the model of the $\nu = 2/3$ state as a 1/3 state of holes on top of a vacuum of a fully filled $\nu = 1$ state, the filling factor must increase from $\nu = 2/3$ in the bulk to $\nu = 1$ at the edge before decreasing to zero. The re-entrant density profile is enabled by the combination of the confining potential with the electrostatic interaction between electrons in the vacuum $\nu = 1$ state and holes that form the inner -1/3 state \cite{Macdonald1990, Macdonald1991, Meir1993}. This results in an inner counterpropagating edge state and outer downstream edge state. In this case, the inner upstream edge state and outer downstream edge state on one edge of a Hall bar can have different chemical potentials because they are emitted from different Ohmic contacts. The situation is shown schematically in Fig. \ref{nu=2_3_cartoons}a. Assuming the edge modes have a small enough separation to allow charge transfer between them, the counterflowing edge modes will equilibrate \cite{Rosenow2018, Protopopov2017, Spanslatt2021} and reach the same chemical potential. Based on the well-established experimental properties of quantum Hall states in bulk samples, the equilibrium chemical potential of each edge should be the potential of the contact from which the downstream, higher-conductance edge state emanated from. This leads to a potential difference $V_{Hall} = \frac{\mu _{S}-\mu_{D}}{e}$, and a two-terminal conductance $G = \frac{2}{3} \frac{e^2}{h}$, consistent with the observed bulk conductance properties of the 2/3 state. A quantum point contact which causes backscattering will take the edge states out of equilibrium and reduce the conductance of the device. In Supp. Section 3 we analyze the situation of a bulk Hall bar and of a QPC when a counter-propagating edge mode is present using a simple model which includes incoherent inter-edge scattering along the edge of the Hall bar (similar models were discussed in previous works \cite{Rosenow2018, Lin2019}. For a Hall bar, there will be equilibration near each Ohmic contact (shown with dashed line in Fig. \ref{nu=2_3_cartoons}a) so that the counterpropagating edge modes come to the same chemical potential; $\mu$ versus position is plotted qualitatively in Fig. \ref{nu=2_3_cartoons}b. For a QPC which fully reflects the inner mode and transmits the outer mode (Fig. \ref{nu=2_3_cartoons}c and d), there will be additional equilibration at the QPC, and the conductance will be reduced to $G = \frac{1}{2}\frac{e^2}{h}$, matching the value of the intermediate plateau in Fig. \ref{nu=2_3_QPCs}c.

An assumption for applying this model to our device is that the equilibration length $l_0$ is much smaller than the length from the QPC to the source and drain Ohmic contacts. In our devices, this length is very long at $\approx 500$ $\mu$m, making full equilibration likely. Equilibration at $\nu = 2/3$ has been studied previously in GaAs \cite{Lin2019, Maiti2020} and graphene \cite{Kumar2022}.



The requirement of finite bias to observe the $G = 0.5 \frac{e^2}{h}$ plateau suggest that at zero bias, there is a wide range of gate voltage over which the outer mode is partially backscattered, so that there is no region where the outer mode is fully transmitted while the inner is fully reflected. Finite bias tends to reduce backscattering in Luttinger liquids \cite{Chamon1997}, so the emergence of the quantized intermediate plateau at $V_{SD}= 50$ $\mu$V may be due to the suppression of backscattering for the outer mode in this regime. Full reflection of the inner mode would be achieved by moving the edges together via more negative gate voltage until the outer strips of $\nu = 1$ fluid merge in the middle of the QPC so that the interface between $\nu = 1$ and $\nu = 2/3$ is absent inside the QPC; thus the inner mode cannot propagate through. The lack of a fully quantized plateau at zero bias might indicate that the electrostatics in the QPC do not enable a region reaching full $\nu = 1$ filling inside the QPC region once the inner mode is fully reflected, but rather the filling is somewhat lower, resulting in a small amount of simultaneous backscattering of the outer mode and $G< 0.5 \frac{e^2}{h}$ as observed in Fig. \ref{nu=2_3_QPCs}c after the initial sharp drop in conductance. Disorder-mediated tunneling may also be a contributing factor \cite{Ensslin2014}. A prediction of Luttinger liquid theory is that when $V_{SD}$ is increased, transmission of the edge state through a barrier will increase until it reaches full transmission, which agrees well with our observation that after a small $V_{SD}$ is applied the conductance increases to the half-integer quantized value indicating full transmission of the outer edge state, and then remains nearly constant \cite{Wen1991_2}. Although integer quantum Hall edge states are generally expected to exhibit Fermi liquid rather than Luttinger liquid behavior based on theory \cite{Wen1991_2}, when an integer edge state lies parallel to a co-propagating or counter-propagating edge state this may induce Luttinger liquid behavior \cite{Hashisaka2018, Fujisawa2022}. Additionally, non-linear conductance suggestive of Luttinger liquid behavior has been observed for the $\nu =1$ integer quantum Hall state in QPCs \cite{Roddaro2005}, suggesting that in real systems even isolated integer edge states may have an interaction parameter that is not exactly 1 and may thus diverge from ideal Fermi liquid behavior.

To our knowledge half integer quantized conductance in a QPC at $\nu = 2/3$ has not previously been reported. On the other hand, in a few experiments a conductance plateau with $G = 1.5 \frac{e^2}{h}$ has been observed in QPCs at $\nu = 5/3$ \cite{Fu2019, Hayafuchi2022, Yan2022}, which is the equivalent of the $\nu = 2/3$ state with opposite spin. We also observe an intermediate plateau at $G = 1.5 \frac{e^2}{h}$ at $\nu = 5/3$ in both QPCs in our device (Supp. Fig. 4a and Supp. Section 4), which suggests that an edge structure occurs at $\nu =5/3$ similar to the one at $\nu = 2/3$ in our device. On the other hand, at the particle-like fractional quantum Hall states $\nu = 4/3$ and $\nu = 2/5$, we do not observe anomalous intermediate plateaus that would imply a re-entrant density profile, which is consistent with the expectation that the $\nu=1$ strip and re-entrant density profile occur only for hole-like fractional quantum Hall states. This distinguishes our work from Ref. \cite{Yan2022}, in which anomalous plateaus were observed even for particle-like states such as $\nu = 4/3$. See Supp. Section 5 for a comparison of our experiment to these previous works.

Additionally, evidence for partially unequilibrated counter-propagating charge transport at short distances has been observed at $\nu = 2/3$, but only in the spin-unpolarized state \cite{Heiblum2019}. Due to the fact that our 2DEG has relatively high density and uses a wide, symmetric quantum well, the 2/3 state is likely polarized in our system \cite{Eisenstein1990, Vanovsky2013}. Numerical studies in the composite fermion picture predict counterpropagating modes for both the spin polarized and unpolarized $\nu = 2/3$ states \cite{Jain2012}. The interplay of spin polarization and confining potential at $\nu = 2/3$ requires more investigation.

\begin{figure}[h]
\def\ffile{Non_SW_v4.pdf}
\centering
\includegraphics[width=1.0\linewidth]{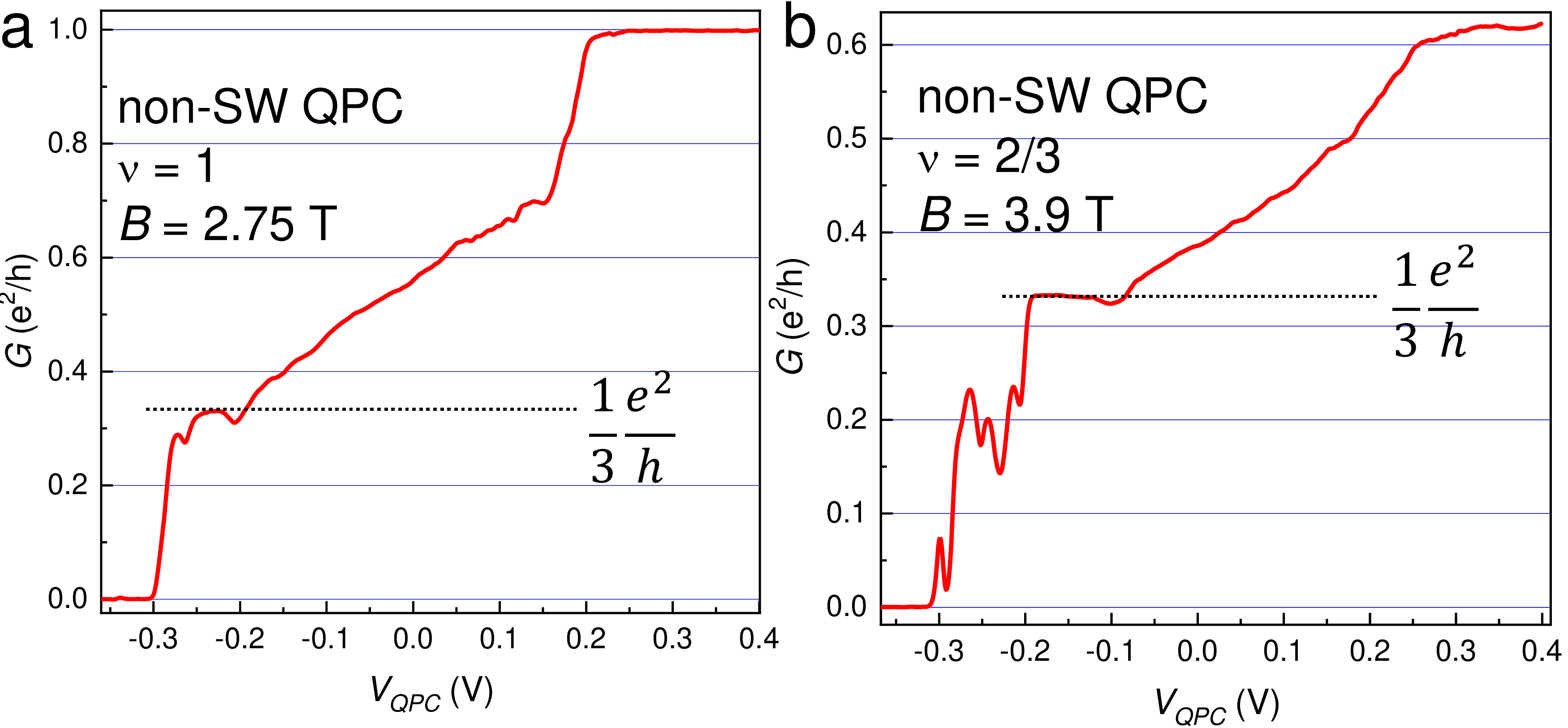}
\caption{\label{Non_SW}a) Gate sweep at $\nu = 1$ for the non-SW QPC. A quasi-plateau at $G = \frac{1}{3} \frac{e^2}{h}$ is visible. b) Gate sweep at $\nu = 2/3$ for the non-SW QPC. A clear intermediate plateau at $G = \frac{1}{3} \frac{e^2}{h}$ is visible.  }
\end{figure}

In order to confirm that the use of the screening well heterostructure enables the observation of the $G = \frac{1}{2} \frac{e^2}{h}$ plateau, we have also measured a QPC fabricated on a standard non-screening well heterostructure. The standard heterostructure uses a single GaAs/AlGaAs interface rather than a quantum well to confine the 2DEG. Fig. \ref{Non_SW}a shows the standard structure (layer stack shown in Supp. Fig. 1b). Fig. \ref{Non_SW}a shows conductance versus gate voltage at $\nu = 1$ (note that the range of gate voltage is smaller because the gate potential is not screened by the screening layers). In contrast to the wide plateau and sharp pinchoff in the screening well QPCs, this standard structure QPC shows a narrow primary plateau with $G = \frac{e^2}{h}$ and a much more gradual decrease in conductance. At $\nu = 1$ there also appears to be an intermediate quasi-plateau at $G \approx \frac{1}{3} \frac{e^2}{h}$, consistent with the phenomenon of edge reconstruction reported in \cite{Heiblum2019-2} and theoretical analysis for the case of soft confinement \cite{Khanna2021}. Fig. \ref{Non_SW}b shows conductance versus gate voltage at $\nu = 2/3$, where a clear intermediate plateau at $G = \frac{1}{3} \frac{e^2}{h}$ is present, consistent with previous results using non-screening well heterostructures \cite{Heiblum2017, Heiblum2019-2}. These results support the theoretical picture that a soft confining potential (as usually realized in standard GaAs/AlGaAs heterostructures not utilizing screening layers) results in a reconstructed edge at $\nu = 2/3$ with an outer strip of filling $\nu = \frac{1}{3}$ and two downstream $\frac{1}{3}\frac{e^2}{h}$ charge modes, as described by the Meir edge structure \cite{Meir1993, Meir2013}. The absence of this $G = \frac{1}{3} \frac{e^2}{h}$ plateau in the QPCs with sharp confinement indicates that the outer $\nu = \frac{1}{3}$ strip does not form, and thus a different edge structure is present.


Our experiments clearly show the importance of confining potential on edge structure in the quantum Hall regime. The lack of edge reconstruction may explain why it has been possible to observe Aharonov-Bohm interference interference at quantum Hall states such as $\nu = 1$ and $\nu = 1/3$ in devices using the screening well heterostructure \cite{Nakamura2019, Nakamura2020, Nakamura2022}, but not in devices using standard non-SW heterostructures \cite{Heiblum2019-2, Heiblum2022}.

\section{Acknowledgements}
This work is supported by the U.S. Department of Energy, Office of Science, Office of Basic Energy Sciences, under award number DE-SC0020138.

\end{document}


\renewcommand{\figurename}{SUPP. FIG.}

\title{Supplementary Material for ``Half-integer conductance plateau at the $\nu = 2/3$ fractional quantum Hall state in a quantum point contact''}

\maketitle

\section{Supplementary Section 1: Screening Well Heterostructure}

\begin{figure}[h]
\def\ffile{SW_Structure_v2.pdf}
\centering
\includegraphics[width=1\linewidth]{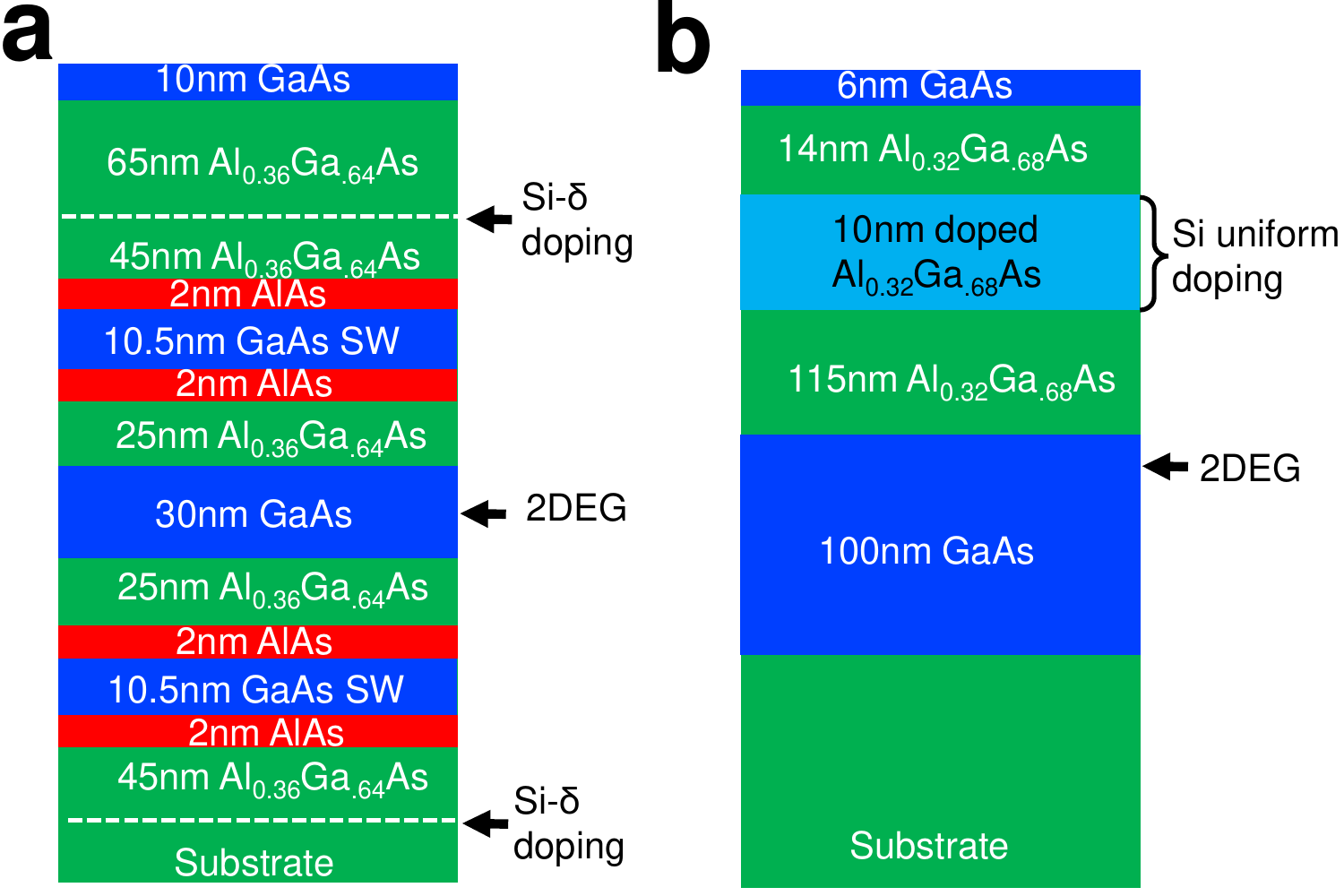}
\caption{\label{SW} a)Layer stack for screening well heterostructure. b)Layer stack for the non-screening well heterostructure. }
\end{figure}

The main device used in this work uses the screening well heterostructure shown in Supp. Fig. \ref{SW}. The structure is grown by molecular beam epitaxy. Similar structures were used in previous works \cite{Nakamura2019, Nakamura2020, Nakamura2022}. The device includes additional gates on the surface and on the back side of the chip which go around the Ohmic contacts, and these gates are biased to just deplete the electrons from the screening wells so that the Ohmic contacts are disconnected from the screening wells and no parallel conduction occurs in the magnetotransport.

\begin{figure}[h]
\def\ffile{B_0_QPCs.pdf}
\centering
\includegraphics[width=1\linewidth]{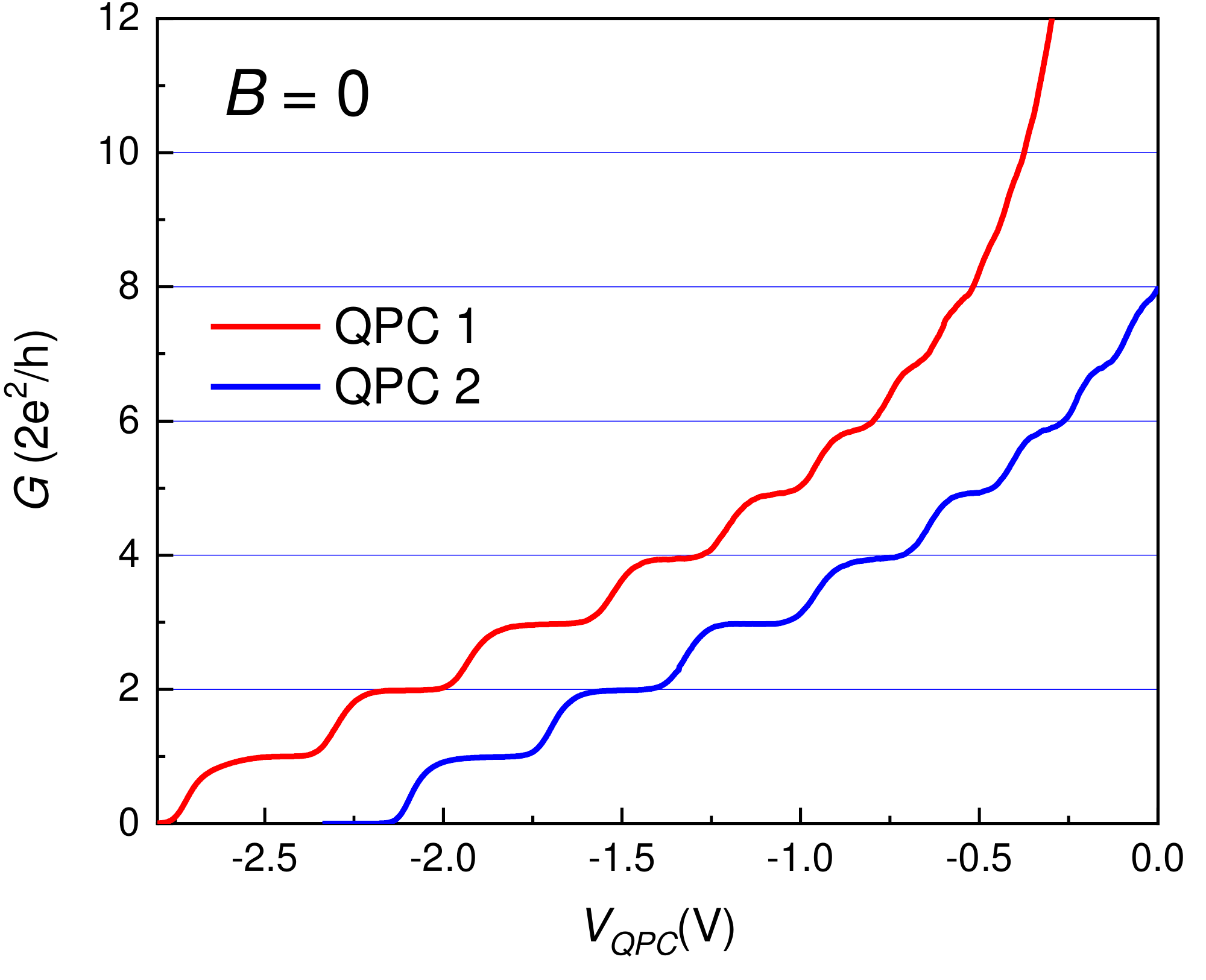}
\caption{\label{B_0} QPC conductance versus gate voltage at zero magnetic field. Conductance plateaus occur with quantized values in increments of $2 \frac{e^2}{h}$ indicating transmission of conductance modes. }
\end{figure}

\section{Supplementary Section 2: Replication in a second set of QPCs}

\begin{figure}[h]
\def\ffile{Supp_Device2.pdf}
\centering
\includegraphics[width=1\linewidth]{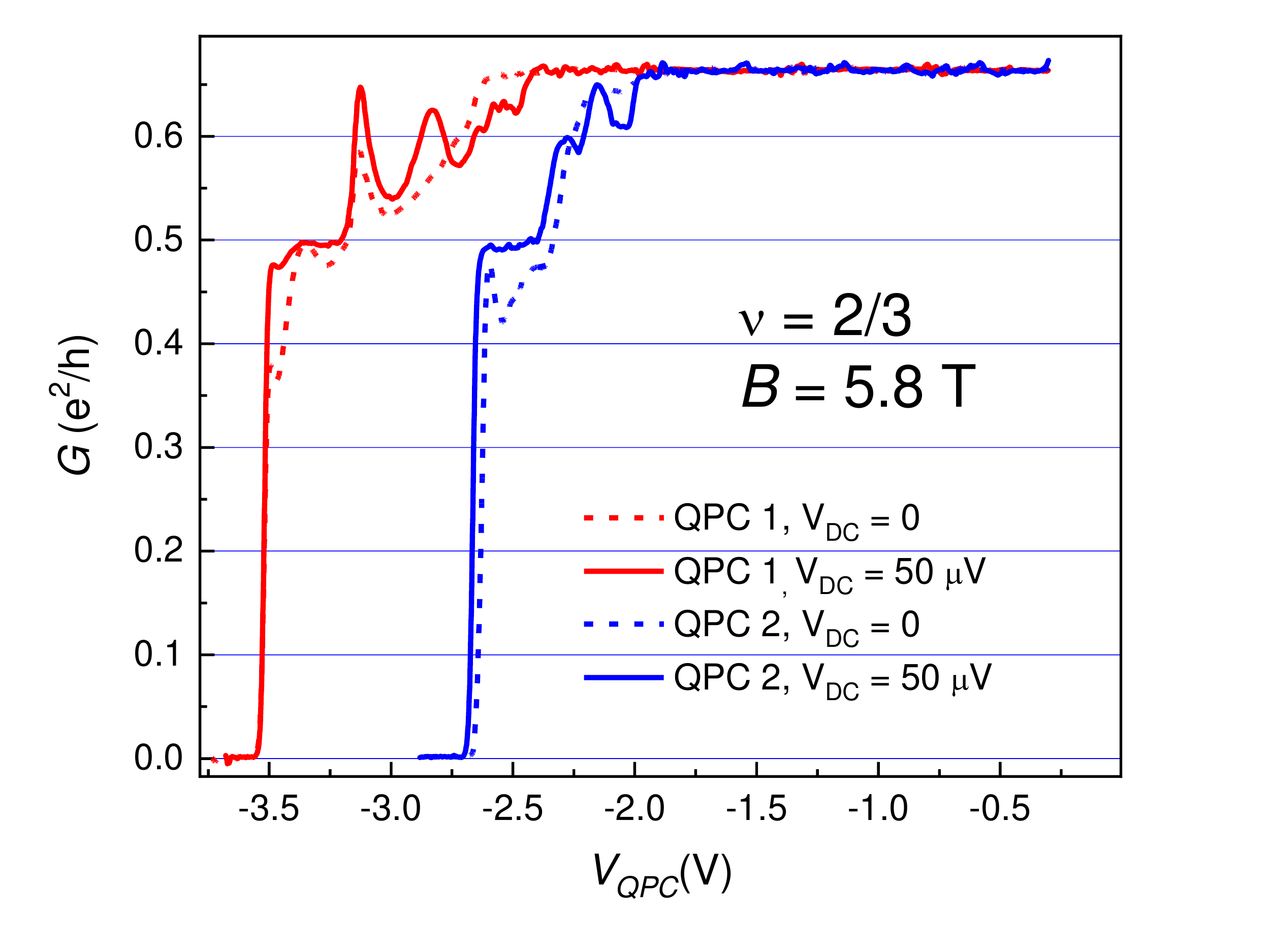}
\caption{\label{repeat} QPC sweeps in a two QPCs on a second device with zero applied DC bias (dashed lines) and with +50 $\mu$V (solid lines). Consistent with the results from the device discussed in the main text, there is an intermediate plateau in at $G = \frac{1}{2} \frac{e^2}{h}$ when the bias is applied. }
\end{figure}

We have measured QPC conductance in a second set of QPCs fabricated on a separate chip on the same screening well wafer at $\nu = 2/3$. Similar to the QPCs discussed in the main text, these QPCs both exhibit a conductance plateau at $G = 0.5 \frac{e^2}{h}$ as shown in Supp. Fig. \ref{repeat}.

\section{Supplementary Section 3: Edge model}

The conductance of a pair of counterflowing edge modes can be treated in a simple model where the current carried by each is equal to the mode's conductance $G_i$ times its local chemical potential $\mu_i (x)$ at its position along the edge (similar models were discussed in Refs. \cite{Rosenow2018, Lin2019}), $I_1 (x)=\mu_1(x)G_1$ and $I_2(x)=\mu_2(x)G_2$. Equilibration (which has been studied theoretically \cite{Rosenow2018, Protopopov2017, Spanslatt2021})  is taken into account by allowing charge to be exchanged between the two modes and assuming a tunneling current per unit length between mode 1 and mode 2 $dI_{12}/dx$ proportional to the chemical potential difference:

\begin{equation}
\label{dI}
    d I_{12}=\frac{e^2}{h}\frac{\mu_2-\mu_1}{l_0}dx
\end{equation}

Here $l_0$ is the characteristic equilibration length between the two edge modes. Conservation of current requires $I = \mu_1 (x) G_1-\mu_2 (x) G_2 = $ constant. Eqn. \ref{dI} can be used to obtain differential equations for $\mu_1$ and $\mu_2$: 

\begin{equation}
    \label{dmu1}
    \frac{d \mu_1}{dx}=\frac{(\mu_2-\mu_1)G_0}{G_1 l_0}
\end{equation}

\begin{equation}
    \label{dmu2}
    \frac{d \mu_2}{dx}=\frac{(\mu_1-\mu_2)G_0}{G_2 l_0}
\end{equation}

$G_0 \equiv \frac{e^2}{h}$ is the conductance quantum. Considering the bottom edge of the Hall bar in main text Fig. 3b where the inner edge emanates from the source contact, $\mu_1 (0)=\mu_S$, and the outer edge from the drain contact, $\mu_2 (L)=\mu_D$, the differential equations are straightforward to solve. They yield:

\begin{equation}
\label{mu1x}
\mu_1(x) = \frac{I}{G_1-G_2}+\left(\mu_S-\frac{I}{G_1-G_2} \right)e^{-\alpha x}
\end{equation}

\begin{equation}
    \label{mu2x}
    \mu_2(x) = \frac{I}{G_1-G_2}+\frac{G_1}{G_2}\left(\mu_S-\frac{I}{G_1-G_2} \right)e^{-\alpha x}
\end{equation}

\begin{equation}
    \label{Ix}
    I = \frac{(\mu_D-\mu_S \frac{G_1}{G_2}e^{-\alpha L})(G_1-G_2)}{1-\frac{G_1}{G_2}e^{-\alpha L}}
\end{equation}

Here $\alpha\equiv \frac{G_0(G_2-G_1)}{l_0G_1G_2}$, and $L$ is the length of the edge of the Hall bar. For large $L$ (much larger than the characteristic equilibration length $l_0$) $I\approx \mu_D (G_1-G_2)$, reflecting the fact that the edge states equilibrate at the chemical potential of the contact that the higher conductance outer edge state emanates from. The chemical potentials of the inner and outer edge states, $\mu_1$ and $\mu_2$, are qualitatively sketched as a function of length along the Hall bar in main text Fig. 3c. An equivalent relationship will hold on the top edge of the Hall bar so that the total current $I_{total} = (\mu_S-\mu_D)(G_2-G_1) = \frac{2}{3}\frac{e^2}{h}(\mu_S-\mu_D)$, thus recovering the two-terminal conductance $\frac{2}{3}\frac{e^2}{h}$ observed in macroscopic quantum Hall samples at $\nu = 2/3$. Furthermore, in this limit $\mu_2 (0) = \mu_D + \frac{G_1}{G_2}(\mu_S-\mu_D)$ and $\mu_1 (L)=\mu_D$.

This solution can be applied to the case of a QPC that fully transmits the outer mode and fully reflects the inner mode, as illustrated in main text Fig. 3d. The solutions to the differential equations can be applied to each quadrant of the Hall bar, allowing the chemical potential $\mu_{QPC}^{BL}$ on the bottom left edge of the QPC and $\mu_{QPC}^{UR}$ at the upper right edge of the QPC to be solved (note that on the top left and bottom right edges the inner and outer modes will not be at equilibrium due to the backscattering of the QPC, so there will be additional equilibration at those positions, as indicated by the white dashed lines). Applying the solutions to the differential equations in the limit $L>>l_0$ gives the equations:

\begin{equation}
    \label{muBL}
    \mu_{QPC}^{BL}=\mu_{QPC}^{UR}\frac{G_1}{G_2}+\mu_D \left(1-\frac{G_1}{G_2}\right)
\end{equation}

\begin{equation}
    \label{muUR}
    \mu_{QPC}^{UR}=\mu_{QPC}^{BL}\frac{G_1}{G_2}+\mu_S \left(1-\frac{G_1}{G_2} \right)
\end{equation}

Solving yields $\mu_{QPC}^{BL}=\frac{\mu_D G_2+\mu_S G_1}{G_2+G_1}$ and $\mu_{QPC}^{UR}=\frac{\mu_S G_2+\mu_D G_1}{G_2+G_1}$. This results in a total current: 

\begin{equation}
    \label{Iqpc}
    I_{QPC}=(\mu_S-\mu_D)\frac{G_2-G_1}{1+\frac{G_1}{G_2}}
\end{equation}

For the case of $G_1 = \frac{1}{3} \frac{e^2}{h}$ and $G_2=\frac{e^2}{h}$ (as expected for the MacDonald edge structure), the result is $G_{QPC} = \frac{1}{2} \frac{e^2}{h}$, matching the value of the intermediate plateaus observed in main text Fig. 2b for $V_{SD}$ = 50 $\mu$V. The chemical potential on the bottom edge in this situation is sketched in main text Fig. 3e. An abrupt discontinuity occurs for $\mu_1$ in the middle of the Hallbar at the position of the QPC since at that point the inner mode is fully reflected.

\section{Supplementary Section 4: QPC sweeps at other filling factors}
 \begin{figure}[h]
\def\ffile{Other_QPC_2.pdf}
\centering
\includegraphics[width=1.0\linewidth]{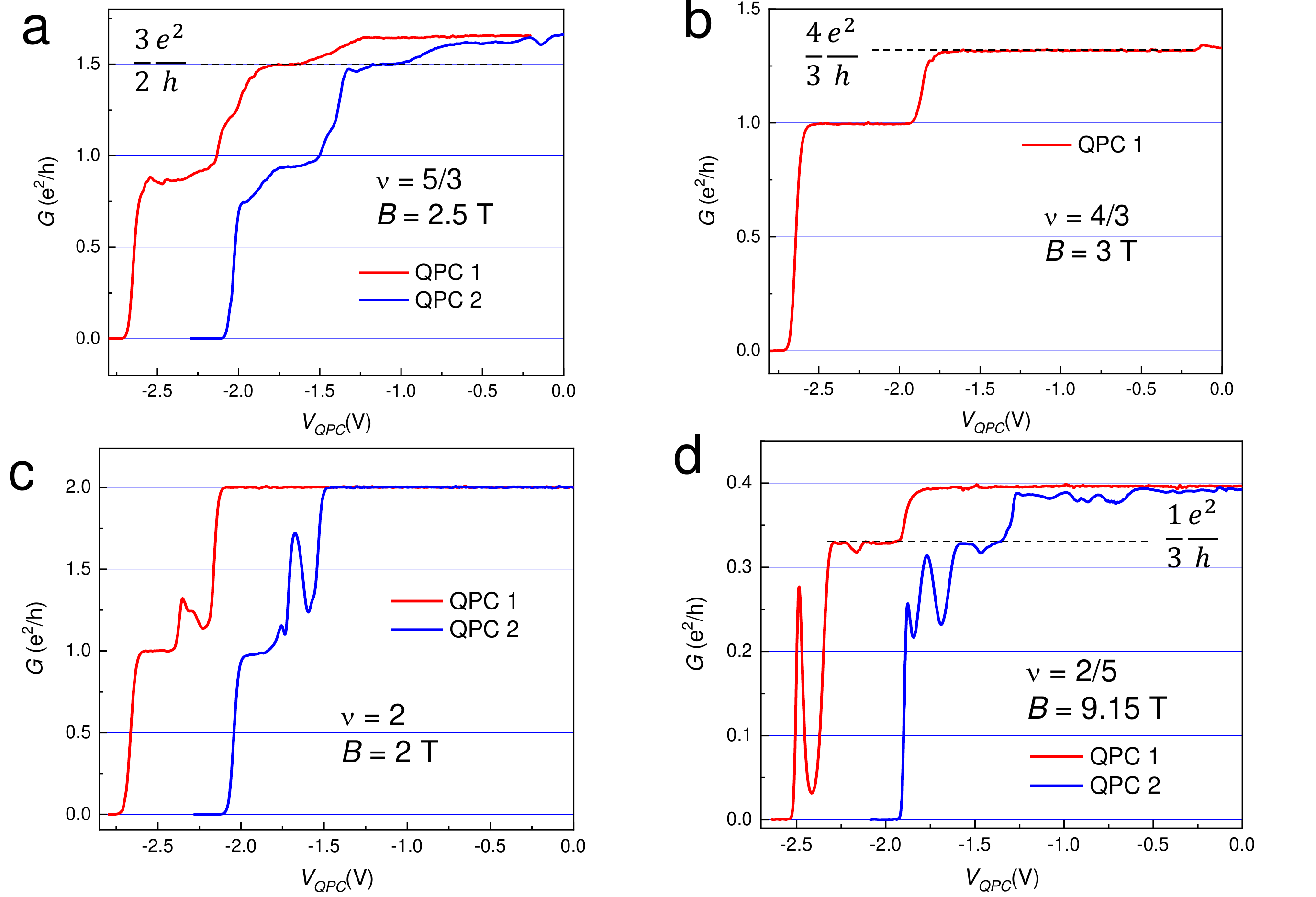}
\caption{\label{Other_QPC}a) QPC sweeps at $\nu = 5/3$. As in the previous plots QPC 2 is shifted by 0.5 V. b) QPC 1 conductance versus gate voltage at $\nu = 4/3$ (we did not perform this measurement for QPC 2 at $\nu = 4/3$. c) QPC sweeps at $\nu = 2$ d) QPC sweeps at $\nu = 2/5$. }
\end{figure}

Similar to the previous experiments \cite{Fu2019, Hayafuchi2022}, we also observe an intermediate plateau at $G = 1.5 \frac{e^2}{h}$ at $\nu = 5/3$ in both QPCs (Supp. Fig. \ref{Other_QPC}a), which suggests that a similar edge structure occurs at $\nu =5/3$ to the one at $\nu = 2/3$ in our device. Notably, this plateau does not require finite DC bias to be observed, and instead occurs at $V_{SD}= 0$. Surprisingly, at $\nu = 5/3$ there is not an intermediate plateau at $G = \frac{e^2}{h}$ which would be expected to occur when only the outermost edge mode is fully transmitted, even though there are $G = \frac{e^2}{h}$ intermediate plateaus at the nearby states $\nu = 4/3$ and $\nu = 5/3$, shown in Supp. Fig. \ref{Other_QPC}b and c. While there does appear to be a rough second intermediate plateau at $\nu = 5/3$, it occurs at conductance somewhat less than $\frac{e^2}{h}$, and the value is somewhat different for QPC 1 and QPC 2. There may non-trivial effects of spin leading to greater scattering between inner and outer edge states than would be naively expected. 

In Supp. Fig. \ref{Other_QPC}b, c, and d we show QPC sweeps at $\nu = 4/3$, $\nu = 2$, and $\nu = 2/5$. Notably, at $\nu = 4/3$ there is a primary plateau at $G = \frac{4}{3} \frac{e^2}{h}$ and an intermediate plateau at $G = \frac{e^2}{h}$, in line with expectations of an inner downstream 1/3 edge mode and outer integer mode. At $\nu = 2/5$, there is a primary plateau at $G = \frac{2}{5} \frac{e^2}{h}$ and intermediate plateau at $G = \frac{1}{3} \frac{e^2}{h}$, consistent with expectations of an inner downstream 1/15 mode (consisting of $e/5$ quasiparticles) and an outer 1/3 mode. The fact that these particle-like fractional quantum Hall states exhibit conductance plateaus quantized to conventional fractions (in contrast to the anomalous $G = \frac{1}{2} \frac{e^2}{h}$ plateau observed at the hole-like $\nu=2/3$ state) is consistent with the expectation that only hole-like states have the re-entrant density profile with a strip of $\nu = 1$ filling at the outer edge.

\section{Supplementary Section 5: Comparison to previous observations of a $G = \frac{3}{2}$}

In a few previous experiments a conductance plateau with $G = 1.5 \frac{e^2}{h}$ has been observed in QPCs at $\nu = 5/3$ \cite{Fu2019, Hayafuchi2022, Yan2022}, which is the equivalent of the $\nu = 2/3$ state with opposite spin. It is noteworthy that these experiments used techniques likely to increase the density of the 2DEG in the QPC. In Ref. \cite{Fu2019} and Ref. \cite{Yan2022}, doping well heterostructures were used, and gate biases were applied at an elevated temperature of 4 K and 6 K, allowing electrons in the X-band of the AlAs layers next to the doping wells to equilibrate with the 2DEG and partially screen the gate voltage \cite{Radu2008}. The $G = 1.5 \frac{e^2}{h}$ plateau was then observed when the gate voltage was increased above the annealed-in value. This unusual decrease in conductance from $\frac{5}{3} \frac{e^2}{h}$ to $\frac{3}{2} \frac{e^2}{h}$ suggests that the half-integer plateau occurred when the density in the center of the QPC was increased above the bulk density due to increasing of the gate voltage after annealing. In Ref. \cite{Hayafuchi2022}, an additional gate in the middle of the device had a positive bias applied, which would be expected to yield higher electron density in the middle of the QPC. A possible explanation for the $G = \frac{3}{2} \frac{e^2}{h}$ plateau is that the techniques of Refs. \cite{Fu2019, Hayafuchi2022, Yan2022} forced the density in the center of the QPC to be higher than the bulk by their external gate voltages, thus yielding $\nu = 2$ in the QPC, which with full equilibration of the additional edge modes in the QPC results in quantized conductance $G = \frac{3}{2} \frac{e^2}{h}$. Similar to the QPC gate voltage dependence in Ref. \cite{Yan2022}, Ref. \cite{Hayafuchi2022} shows a non-monotonic dependence on the center gate voltage, supporting the possibility that the 2DEG density in the middle of the QPC is being forced above the bulk value by the external gate potential. A notable difference in our experiment compared to these previous works is that in our device, the conductance increases nearly monotonically with gate voltage, with the primary $G = \frac{2}{3} \frac{e^2}{h}$ plateau occurring at higher gate voltage compared the $G = 0.5 \frac{e^2}{h}$ plateau, suggesting that the lower-conductance intermediate plateau does not occur because the externally applied gate voltages have forced the density in the QPC to be higher, but because the intrinsic interactions of the electrons favor the formation of a $\nu = 1$ strip and re-entrant density profile at the edge.

Similar to the previous experiments \cite{Fu2019, Hayafuchi2022, Yan2022}, we also observe an intermediate plateau at $G = 1.5 \frac{e^2}{h}$ at $\nu = 5/3$ in both QPCs as shown in Supp. Fig. \ref{Other_QPC}c, which suggests that an edge structure occurs at $\nu =5/3$ similar to the one at $\nu = 2/3$ in our device. On the other hand, at the particle-like fractional quantum Hall states $\nu = 4/3$ and $\nu = 2/5$, we do not observe anomalous intermediate plateaus that would imply a re-entrant density profile, which is consistent with the expectation that the $\nu=1$ strip and re-entrant density profile occur only for hole-like fractional quantum Hall states. This further distinguishes our work from Ref. \cite{Yan2022}, in which anomalous plateaus were observed even for particle-like states such as $\nu = 4/3$.

\section{Supplementary Material References}